\documentstyle[sprocl]{article}

\def\Journal#1#2#3#4{{#1} {\bf #2}, #3 (#4)}

\def\NPB{{\em Nucl. Phys.} B}
\def\PLB{{\em Phys. Lett.}  B}
\def\PRL{\em Phys. Rev. Lett.}
\def\PRD{{\em Phys. Rev.} D}

\def\be{\begin{equation}}
\def\ee{\end{equation}}
\def\bea{\begin{eqnarray}}
\def\eea{\end{eqnarray}}

\begin{document}

\title{Black holes with toroidal, cylindrical and planar horizons in 
anti-de Sitter spacetimes in general relativity and their properties
\footnote{ $\,$in$\,${\it Astronomy and Astrophysics: Recent Developments}, 
Proceedings of the 10$^{\rm th}$ Portuguese 
Meeting in Astronomy and
Astrophysics, Lisbon, July 2000, eds J. P. S. Lemos, A. Mour\~ao,
L. Teodoro, R. Ugoccioni (World Scientific, 2001).}
}

\author{Jos\'e P. S. Lemos}

\address{Centro Multidisciplinar de Astrof\'{\i}sica - CENTRA, 
Departamento de F\'{\i}sica, Instituto Superior T\'ecnico,
\\ Av. Rovisco Pais 1, 1049-001, Lisbon, Portugal\\
E-Mail: lemos@kelvin.ist.utl.pt}

\maketitle\abstracts{ We review the toroidal, cylindrical and planar
black hole solutions in anti-de Sitter spacetimes and present their
properties.}

\section{Introduction}

Black holes have been objects of astrophysical interest after it was
shown that they are the inevitable outcome of complete gravitational
collapse of a massive star or a cluster of stars. They are likely to
power the spectacular phenomena seen in x-ray emitting star accretion 
disks, in active galactic nuclei and quasars, and should also inhabit, in
a quiescent state, the center of normal galaxies, such as our own.

Black holes appear naturally as
exact solutions in general relativity.
Their theoretical properties, such as their stability, the no
hair theorems (stating they are characterized by three parameters only,
the mass, the angular momentum and the electric charge), and the
physics of matter around them, have been established.  These black
holes live in an asymptotically flat spacetime.

Black holes became important, not only to astrophysics, but
also to physics, after the discovery that, due to quantum processes,
they can emit radiation with a specific temperature, the Hawking
temperature. It was also shown that they have a well defined entropy.
Thus, black holes turned to be objects subject to the laws of thermodynamics.

Once they have entered the domain of physics it also became clear that
one should study them not only in asymptotically flat spacetimes, but
also in spacetimes with a positive cosmological constant $\Lambda>0$,
i.e., asymptotically de Sitter spacetimes, and spacetimes with a
negative cosmological constant $\Lambda<0$, i.e., asymptotically
anti-de Sitter spacetimes. From recent astronomical observations, it
seems now that we live in a world with $\Lambda>0$. However, a
spacetime with $\Lambda<0$ is also worth of investigation, since it
allows a consistent physical interpretation. Indeed, one can enlarge
general relativity into a gauged extended supergravity theory, in
which the vacuum state has an energy density given by
$\Lambda=-3g^2/(4\pi G)$, where $g$ is the coupling constant of the
theory, and $G$ is the constant of gravitation. Thus, its vacuum is
described by an anti-de Sitter spacetime. If such types of theories
are correct, it means, at least qualitatively, that anti-de Sitter
should be considered as a symmetric phase of the theory, although it
must have been broken, since we do not live now in a universe with
$\Lambda<0$.  In addition, anti-de Sitter spacetimes have other
interesting features: (i) it is one of the rare gravitational
backgrounds yielding a consistent interaction with massless higher
spins, (ii) it permits to have a consistent theory of strings in any
dimension, not only the critical ones, (iii) it has been conjectured
that it has a direct correspondence with conformal field theory, the 
AdS-CFT conjecture, and
(iv) it allows a good definition for the mass, angular momentum and
charge (a property shared with the asymptotically flat case).

Most important to this review, a whole variety of
classes of black holes exist in asymptotically anti-de Sitter
spacetimes alone, and are amenable to be studied theoretically.
Indeed, one can extended the known solutions of the Kerr-Newman
family, with horizons that are spheres (in the topological sense), to
include a $\Lambda<0$ term, the Kerr-Newman-anti de Sitter family (one
can also extend to include a $\Lambda>0$ term, the Kerr-Newman-de
Sitter family). But, for $\Lambda<0$, there are also black holes with
horizons with topology different from spherical.  The class we are
interested here contains black holes with toroidal, cylindrical or
planar topology found in \cite{lemos2}$^-$\cite{lemoszanchin96b}.  
Many works have
been now done in this and related subjects, which also have many different
connections.

\section{The solution}

The Einstein-Maxwell action is given by 
\begin{equation}
S = \frac{1}{16\pi}\int d^4x \left( R + 6\alpha^2
            - F_{mn}F^{mn}\right) \, ,
            \label{action1}
\end{equation}
where we have put $G=c=1$ and 
the cosmological constant $\Lambda\equiv-3\alpha^2$. 
A rotating solution, which can represent a black hole, may be found
from the equations of motion following from 
(\ref{action1}) \cite{lemoszanchin96}. 
The metric is given by,  
\begin{eqnarray}
ds^{2} &=& - \frac{\Delta}{(\alpha r)^2} (dt-\lambda d\phi)^2 + 
\frac{(\alpha r)^2}{\Delta} dr^2 +
\nonumber\\
&&+(\alpha r)^2 dz^2 + r^2(\alpha^2 \lambda dt - d\phi)^2\, ,
                                                       \label{metric1}
\end{eqnarray} 
and the electric potential by
\begin{equation}
A= -\frac{2Q_E}{\alpha r} (dt-\lambda d\phi) \,.
                                                      \label{potential1}
\end{equation}
The Riemann $R_{klmn}$, Ricci $R_{mn}$, and scalar $R$ curvatures can
be found from (\ref{metric1}). They yield a spacetime singularity at
$r=0$.  The Maxwell tensor $F_{mn}$ can be easily taken from the
Maxwell curvature 2-form $F=F_{mn}dx^m\wedge dx^n$, i.e., $F= dA=
\frac{2Q_E}{\alpha r^2} dr \wedge (dt-\lambda d\phi)$, yielding two
non-zero components for $F_{mn}$, $F_{rt}=2Q_E/(\alpha r^2)$ and
$F_{r\phi}=-2Q_E\lambda/(\alpha r^2)$. A magnetic charge can easily be 
added.

In this solution, the ranges of the time and radial coordinates are
$-\infty<{t}<+\infty$, $0\leq r< +\infty$.  The topology of the two
dimensional space, $t={\rm constant}$ and $r={\rm constant}$, can be
(i) $S^1 \times S^1$ the flat torus $T^2$ model, with $0\leq\alpha z<
2\pi$, $0\leq{\phi}< 2\pi$, (ii) $R\times S^1$, the standard
cylindrical symmetric model, with $-\infty <z<+\infty$,
$0\leq{\phi}< 2\pi$, and (iii) $R^2$, the infinite plane model, with
$-\infty<z<+\infty$, $-\infty<{\phi}<-\infty<$ (this planar solution does not 
rotate). 

The functions and parameters appearing in
(\ref{metric1}) are: $\Delta\equiv 4c^2 -b \alpha r +
\alpha^4 r^4$, $\lambda\equiv\frac{1}{\alpha}
\frac{\sqrt{1-\sqrt{1-\frac{8J^2\alpha^2}{9M^2}}}}
{\sqrt{1+\sqrt{1-\frac{8J^2\alpha^2}{9M^2}}}}$ a specific angular
momentum parameter, $J$ and $M$ being the angular momentum and mass
per unit length of the system, respectively,
$c^2=4Q_E^2(1-\lambda^2\alpha^2)$ with  $Q_E$ 
being the electric charge per unit length 
(for a solution with magnetic charge in addition see section 7), 
and $b\equiv4M\frac{1-\lambda^2\alpha^2}{1+\frac12
\lambda^2\alpha^2}$ is a mass parameter.  
Note that $M=0$ implies $J=0$. For the toroidal case, the
mass $\bar M$, angular momentum $\bar J$, and charge $\bar Q$ are
given in terms of $M$, $J$ and $Q$ by $\bar{M}=\frac{2\pi}{\alpha}
{M}$, $\bar{J}=\frac{2\pi}{\alpha} J$, and
$\bar{Q}=\frac{2\pi}{\alpha} Q$ \cite{lemoszanchin96}.  When $\Delta$
has two roots the solution represents a black hole (in the toroidal
model), a black string also called a cylindrical black hole
 (in the cylindrical model), or a black membrane also called a planar
black hole (in the planar model), when it has one root it represents,
correspondingly, an extreme black hole, an extreme black string, or an
extreme black membrane, and when it has no roots it represents 
a singular closed naked line, a singular naked
straight line, or a singular naked membrane, respectively.

Spatial infinity, $r\rightarrow \infty$, in these solutions has the
same topology of the event horizon. This fact makes the cylindrical
black hole of some astrophysical relevance, in particular, for the
hoop conjecture (see below), whereas the toroidal and planar black
holes seem to be of no or little astrophysical interest.

We note that the spacetime metric as given in  
(\ref{metric1}) does not yield the usual
form of anti-de Sitter cylindrical spacetime at spatial
infinity. However, one can always rescale the coordinates $r$ and $z$
so that the usual form becomes apparent.

The non-rotating $J=0$, uncharged $Q=0$ black hole can be read from 
(\ref{metric1}). It has a simple form
\begin{equation}
ds^2=-\left(\alpha^2r^2-\frac{b}{\alpha^2\, r^2}\right) dt^2 +
\frac{dr^2}{\alpha^2r^2-\frac{b}{\alpha^2\, r^2}}+
\alpha^2 r^2\,dz^2 + r^2\,d\phi^2\,,
                                      \label{metricnonrotuncharged}
\end{equation}
from which the radius of the event horizon is readily taken, 
$r_{\rm h}=\frac{b^{1/3}}{\alpha}$. If we restrict the metric 
(\ref{metricnonrotuncharged}) to the toroidal case then it admits 
an extra parameter, the  metric on 
the flat torus $T^2$ can have a more general description. 
Indeed, the two-manifold 
$t=\,$constant, $r=\,$ constant in  (\ref{metricnonrotuncharged}) 
has a metric given by  $d\sigma^2=d\theta^2+d\phi^2$,
where we have put $\theta=\alpha\, z$. 
One can generalize this into the most general form of a metric on 
the torus, i.e., $d\sigma^2=|\tau|^2d\theta^2+d\phi^2 +2\, ({\rm Re\,}
\tau) \,d\theta \,d\phi$, where $\tau$ is the Teichm\"uler complex 
parameter on the torus  giving classes of 
conformally equivalent tori (Im$\,\tau>0$). This parameter 
has several implications on the structure of the black holes, for 
instance, the size of the event horizon depends on the conformal 
class of the torus \cite{vanzo}, 
$r_{\rm h}= \frac{2}{\pi}\frac{\bar{M}}{\alpha^2\,|{\rm Im}\,\tau|}$.

In terms of the $\lambda^2\alpha^2$ angular-momentum, charge $Q$, and mass $M$
one can distinguish clearly five zones:
(i) $0\leq 
\lambda^2\alpha^2<1-\frac{64Q^6}{27M^4}\left(1+\alpha^2\lambda^2\right)^4$: 
black hole region, 
(ii) $\lambda^2\alpha^2=1-\frac{64Q^6}{27M^4}
\left(1+\alpha^2\lambda^2\right)^4$: extreme black hole line,
(iii) $1-\frac{64Q^6}{27M^4}\left(1+\alpha^2\lambda^2\right)^4
<\lambda^2\alpha^2<1$: naked singularity region,
(iv) $\lambda^2\alpha^2=1$: null singularity line,
(v) $1<\lambda^2\alpha^2\leq2$: pathological black hole region 
(black holes with closed timelike curves).
For $J>\sqrt{\frac{9M}{3\alpha}}$, the specific angular momentum 
parameter $\lambda$ turns complex, and the metric is ill-defined. 
The extreme case is given when $Q$ is connected to $\lambda$ 
through the relation $Q^6=\frac{27}{64}\frac{1-
\lambda^2\alpha^2}{1+\frac12\lambda^2\alpha^2}$. 
For the Penrose diagrams see \cite{lemoszanchin96,lemoszanchin96b}.
In the uncharged case, one has three zones only \cite{lemos1},
(i)   $0\leq\lambda^2\alpha^2<1$:  black hole region,
(ii)  $\lambda^2\alpha^2=1$:  extreme black hole line, and
(iii) $1<\lambda^2\alpha^2\leq2$: naked singularity region.
The metric with electric charge and zero angular momentum was 
also discussed in \cite{huang}.
For the Penrose diagrams see \cite{lemoszanchin96b}.

The black holes as well as the naked singularities can appear through
gravitational collapse, the black holes display quasi-normal modes,
have interesting properties due to their horizon topology, the
non-rotating configurations have thermodynamical properties related to
Reissner-Nordstr\"om black holes, some of these black holes 
are supersymmetric, and there are many 
connections with other black hole solutions, as will be 
shown in the next sections.

\section{Gravitational collapse}

From the work of Oppenheimer and Snyder \cite{oppenheimer} and
Penrose's theorem \cite{penrose65} we know that if general relativity
is correct, then realistic, slightly non-spherical, complete collapse
leads to the formation of a black hole and a singularity. This result
tights strongly the theory of gravitational collapse to the theory of
black holes. In this connection, two important conjectures were
formulated, the cosmic censorship and the hoop conjectures.  The
cosmic censorship conjecture \cite{penrose69} forbids the existence of
naked singularities, singularities not surrounded by an event horizon
of a black hole.  The hoop conjecture \cite{thorne72} states that black holes
form when and only when a mass $M$ gets compacted into a region whose
circumference in every direction is less than its Schwarzschild
circumference $4\pi M$ ($G=c=1$).

Thus, motivated by the cosmic censorship and hoop conjectures, 
it is important to understand,
besides spherical collapse, highly non-spherical collapse. The
collapses of prolate spheroids, such as spindles, or oblate spheroids,
such as pancakes, are not only astrophysically interesting but also
important to a better understanding of these conjectures. Prolate
collapse has been studied in some detail \cite{echeverria}
and it was shown that fully relativistic effects, totally different
from the spherical case, come into play. In addition, cylindrical
systems were used by Thorne to mimic prolate collapse \cite{thorne72}.
Oblate collapse has been mainly studied within Newtonian theory, in
connection with galaxy formation, but relativistic results are also
available.

All these results were applied to spacetimes with zero cosmological 
constant. In the presence of a negative cosmological constant 
one can expect the occurrence of major changes.
Indeed, we showed in section 2 that there are black hole solutions with 
toroidal, cylindrical or planar topology 
if a negative cosmological constant is present. 
Since the importance of the spherical Schwarzschild black hole 
has first come from its role as the final state of complete gravitational 
collapse of a star, it is useful to investigate if these black holes with 
different topology may also emerge from gravitational collapse of some 
matter distribution. Moreover, one would like to test the cosmic censorship 
and hoop conjectures in these cases.

An important feature of spherical collapse onto a Schwarzschild black
hole is that, due to Birkhoff's theorem, spacetime is static outside
the matter and the collapse proceeds without emission of gravitational
waves. The same is true for the collapse of spherical matter in an
anti-de Sitter background.  On the other hand, it is well known that
the collapse of cylindrical systems proceeds with emission of
gravitational waves which creates additional problems in their
modeling.  However, surprisingly, the problem can be solved exactly
\cite{lemoscollapse1} by using, for the exterior spacetime, a modified
Vaidya metric appropriate to the cylindrical collapse (we refer here
to the cylindrical case, only because it is the most interesting for
gravitational collapse and for the above conjectures, although the
results apply equally well to toroidal an planar collapses).  This
modified Vaidya metric describes the gravitational field associated
with a cylindrical flow of unpolarized scalar, neutrino,
electromagnetic or gravitational radiation in the geometrical optics
approximation.  The interior solution, is a modified Friedmann
solution also applicable to cylindrical topology.  Through a smooth
matching at the interface, we find that the flux of waves modeled by
the modified Vaidya metric is an incoming flux, and consequently that
the mass parameter of the collapsing matter grows up to the formation
of the black hole.  By carefully choosing the right amount of incoming
flux one avoids the emission of gravitational waves from the
collapsing matter.  One can then show that that a black hole forms,
and study the physics from the inside and outside points of view. It
has much the same characteristics of the formation of a Schwarzschild
black hole \cite{lemoscollapse1}.

It is also possible to test the cosmic censorship and hoop conjectures
\cite{lemoscollapse2} for collapse in a background with a negative
cosmological constant.  It was found \cite{lemoscollapse2} that (i) in
spherical collapse with negative $\Lambda$, massless naked
singularities form for a sufficiently inhomogeneous collapse,
similarly to the $\Lambda=0$ case, and (ii) cylindrical (planar or
toroidal) collapse with negative $\Lambda$ does not produce naked
singularities, in accordance with the cosmic censorship; instead black
strings (black membranes or black holes, respectively) will form
giving an explicit counter-example to the hoop conjecture.  For
charged collapsing matter naked singularities form, violating the
cosmic censorship \cite{ghosh2000}.

To theoretical astrophysics the interesting black hole is the one with
cylindrical topology although an object with cylindrical topology (and
symmetry) in an anti-de Sitter cylindrical spacetime is an idealized
situation, of course.  However, it is possible that the universe we
live in contains an infinite cosmic string. It is also possible, that
the universe is crossed by an infinite black string.  The solution
with planar topology, a black membrane, can be of some relevance.  On
the other hand, the toroidal topology does not seem to be of immediate
interest to astrophysics.  The reason is that the flat torus is
constructed from multiplying two circles. Since each circle lives in a
two dimensional plane, the flat torus could only be visualized inside a
four dimensional Euclidean (flat) space, not in a three dimensional one like
ours.  However, if the universe has some exotic topology at infinity
it could accommodate a flat torus black hole (see section on the
topology of the horizon below).  A flat torus black hole can also be
of relevance to theories of elementary particles invoking extra
dimensions.

\section{Quasi-normal modes}

Perturbations of stars and black holes reveal much of their intrinsic 
structure. For compact objects such as neutron stars and black holes 
the study of perturbations is closely linked to the gravitational 
wave emission and is therefore relevant to gravitational wave 
astronomy. Much work has been done in asymptotically flat 
spacetimes \cite{kokkotas99}.
It is thus also important to study the properties under perturbations 
of the black holes that live in anti-de Sitter spacetimes and 
test for their stability. 

Some objects that we are used to perturb, for instance a string,
display a whole set of normal modes (with real frequencies) which give
interesting and important properties. For gravitational objects, such
as a black hole, as the system starts to vibrate it emits
gravitational waves and damps down. In this case, due to the emission
of gravitational waves, normal modes do not exist. Instead, there are
quasi-normal modes with frequencies that have real parts (signalling
the vibration of the hole) and imaginary parts (representing the
damping down).  It is, of course, important that the imaginary part of
the frequency yields a damping mode (not a growing one) such that the
whole perturbation is stable. This is the case for the spherical
horizon of the Schwarzschild black hole \cite{kokkotas99}. It is also
the case for perturbations of spherical black holes in anti-de Sitter
spacetimes, see \cite{hubeny} and \cite{cardoso1} for references, as
well as for toroidal, cylindrical and planar black holes
\cite{cardoso2}.  The emission of waves from a particle falling down
into an anti-de Sitter black hole can also be considered
\cite{cardoso3}.

\section{The Topology of the Horizon}

The topology of the horizon of the black holes we have been presenting
can be toroidal, cylindrical and planar. However, it was shown by
Hawking \cite{hawkingellis} as a theorem, that under certain
reasonable physical conditions (spacetime is asymptotically flat and
globally hyperbolic, and the dominant energy condition holds), the
topology of the event horizon is spherical. This result was hinted by
the topology of the known black holes, the Schwarzschild and Kerr
black holes, and their charged generalizations, where the shape of the
event horizon is spherical (or easily molded into a spherical
shape). Thus it came as a surprise, that the black holes we have
discussed found in
\cite{lemos2}$^-$\cite{lemoszanchin96b}, with topologies
other than spherical are solutions of Einstein's equations. The reason
that they are not in accord with Hawking's theorem is that, due to the
presence of a negative cosmological constant, spacetime is neither
asymptotically flat nor globally hyperbolic.  
In order to accommodate these topologies into
general relativity one has to generalize Hawking's theorem. This can be
done by using the topological censorship theorem \cite{friedmann93},
which states that the topology of spacetime cannot be directly probed by
distant observers.  Using this idea, one can generalize the theorem
valid for asymptotically flat spacetimes and show that there is a
relationship between the topology of the horizon and the topology at
infinity \cite{galloway99} (see \cite{tedjac} also).  It is then
possible to have toroidal, cylindrical or planar horizons in a
toroidal, cylindrical or planar spacetimes at infinity.

\section{Thermodynamic properties}

We now turn to quantum aspects. We first study the
thermodynamical properties of these black holes and compare them to
their spherically symmetric counterparts.  One direct approach 
to thermodynamics is through 
the path-integral approach. Its application 
to the thermodynamics of black holes was originally developed by
Hawking  et al. \cite{hartle}$^-$\cite{hawking2}. In this approach
the thermodynamical partition function is computed from the
path-integral in the saddle-point approximation, allowing one to obtain 
the thermodynamical laws for black holes.  In the path-integral approach
we can use the three different ensembles: microcanonical, canonical
and grand canonical. Due to difficulties related to stability of the
black hole in the canonical ensemble, the microcanonical ensemble was
originally considered \cite{hawking2}. However, further developments
by York et al. \cite{york,grqcbrown} allowed to define the
canonical ensemble.  Effectively, by carefully defining the boundary
conditions, one can obtain the partition function of a black hole in
thermodynamical equilibrium. This approach was further developed to
include other ensembles, and to study charged black holes in the grand
canonical ensemble and black holes in asymptotically anti-de~Sitter
spacetimes \cite{brillloukopeldan,pecalemos1}. In York's formalism the black
hole is enclosed in a cavity with a finite radius. The boundary
conditions are defined according to the thermodynamical ensemble under
study.  Given the boundary conditions and imposing the appropriate
constraints, one can compute a reduced action suitable for doing black
hole thermodynamics. Evaluating this reduced action at its stable
stationary point one obtains the corresponding classical action, which
is related to a thermodynamical potential.  In the canonical ensemble
this thermodynamical potential corresponds to the Helmholtz free
energy, while for the grand canonical ensemble the thermodynamical
potential is the grand canonical potential. From the thermodynamical
potential one can compute all the relevant thermodynamical quantities
and relations.

In order to compare the thermodynamics of charged black holes with
spherical horizons \cite{pecalemos1} and charged black holes with
toroidal horizons \cite{pecalemos2} we have used York's formalism in
the grand canonical ensemble. An important quantity is the entropy,
which for toroidal black holes is given by $S = \pi^2 r_+^2\,$.  where
$r_+$ is the horizon radius of the charged black hole.  Since $\pi^2
\, r_+^2= A_+/4$, where $A_+$ is the area of the event horizon, we
have $S=\frac{A_+}{4}$. This is the usual Hawking-Bekenstein entropy,
which means this law is still valid for black holes with toroidal
symmetry.  Another important quantity is the heat capacity $C$ of the
black hole. For a simple case, when the boundary of the cavity goes to
infinity one gets \cite{pecalemos2}, $C = 2 \, \pi^2 \, r_+^2 \ \left(
1- \frac{2\, Q_E^2}{Q_E^2+3 \, \pi^2 \, \alpha^2 \, r_+^4}\right) \,$.
One can compute more general cases, for instance when the boundary of
the cavity is held at a finite value $r_B$ 
\cite{brillloukopeldan,pecalemos2}.  The
heat capacity in this case is always positive which means
these solutions are all stable.  Furthermore we find that in the grand
canonical ensemble, with temperature and electrostatic potential fixed
at the boundary, there is a black hole solution that is globally
stable, which means it dominates the grand partition function. These
results are generally different from the results obtained for the
spherical counterpart of this black hole, the
Reissner-Nordstr\"om-anti-de~Sitter black hole, for which there are
one or two solutions, that can be stable or unstable, and do not
necessarily dominate the grand partition function
\cite{pecalemos1}. Thus, contrary to the
Reissner-Nordstr\"om-anti de~Sitter black hole, for the toroidal black
hole no phase transition was found. 

For other studies of thermodynamic properties of these black holes 
see \cite{vanzo,emparan}, and for the subtle issue of the entropy of 
extreme black holes see \cite{ghoshmitra}.

\section{Supersymmetries}

To go beyond the astrophysical properties of the black holes, 
such as the issue of gravitational collapse, and try 
to understand a black hole as a particle, a massive elementary 
particle, it is also important to embed these objects in some 
comprehensive 
particle theory, beyond general relativity, such as a supergravity theory.
Black holes in these theories are called supersymmetric black holes. 
There are several motivations to, first, find supersymmetric black
hole solutions and, then, analyze their properties. For instance,
black holes that are supersymmetric, have zero temperature, and in
some cases it is possible to associate them with solitons of the
underlying theory, interpolating between two distinct homogeneous vacua
solutions.  In addition, in certain instances, supersymmetric black
holes are exact solutions of the theory even when quantum corrections
are taken into account.

There are by now many supersymmetric black hole solutions in many
different supergravity theories.  A characteristic of supersymmetric
theories is that there are as many bosons as fermions. We are
interested in a particular theory, called $N=2$ gauged supergravity,
since it yields the Einstein-Maxwell theory with a negative
cosmological constant, when there are no fermions. One can then look
for the black hole solutions we have found in section 2, which remain
the same when one makes the allowed symmetry transformations of the
underlying theory.  These black hole solutions are the supersymmetric black
holes.  Normally, since one is treating a theory of  elementary particles,
one also includes in the solutions magnetic charge in addition to
electric charge. 

The supersymmetry of the spherical solutions, such as the
Schwarzschild anti-de Sitter 
solution, studied in \cite{romans92,kosteleckyperry96}
show that to be supersymmetric the black hole must rotate. Subsequent
studies for other types of black holes (see section 8.1) were performed in
\cite{caldarelliklemm98,alonsoalercaetal00}, with additional features
not present in the spherical case.

For the black hole presented in section 2, one can also study its
supersymmetry properties \cite{lemos3}.  The study of the
integrability conditions has shown that the supersymmetric black hole
has zero electric and magnetic charge, must rotate with extreme
angular velocity and thus has zero temperature.  The corresponding
Killing spinor is $\epsilon= \sqrt{r} \left[\frac12
(1-\gamma_1)\right] \epsilon_0\, ,$ where $\gamma_1$ is a Dirac matrix
in the real representation, and $\epsilon_0$ is a constant Dirac
spinor.  Due to the projection operator $\frac12 (1-\gamma_1)$, this
equation defines two linearly independent Dirac spinors. Thus, this
extreme uncharged black hole has two supersymmetries, the same number
as the background spacetime. There are other non-rotating solutions,
representing charged naked singularities, which are
supersymmetric. For further details see \cite{lemos3}.

\section{Connections to other black holes}

\subsection{Non-isometric toroidal solutions and multi-tori solutions}

The rotating solution (equations
(\ref{metric1})-(\ref{potential1})) can be obtained from the
static toroidal metric by mixing time and angle into a new time
and a new angle. Since angles are periodic, this is not a proper
coordinate transformation, yielding a new solution
globally different from the static one (see \cite{lemos1}). 

A different rotating toroidal metric can be extracted from the general
Petrov type-D solution by choosing the corresponding parameters in an
appropriate way and imposing time-angle symmetry $t \to -t$ and
$\phi\to-\phi$ \cite{klemm}.  This metric cannot be obtained by the
above coordinate mixing, and thus, it implies that flat torus topology
for the $r={\rm constant}$, $t={\rm constant}$ surfaces admits at
least two distinct rotating black hole solutions not isometric to each
other, although when the rotation parameter is zero 
they are isometric.

A toroidal black hole has a smooth orifice
(non-black) in its middle, so to speak. The event horizon is a Riemann 
surface with genus $g=1$. There are other solutions with
more than one orifice through the horizon, i.e., black
hole solutions with higher genus, $g>1$. These other types of topological 
black holes were
found in \cite{aminneborg}$^-$\cite{smithmann} and their properties studied.
Black holes with exotic horizons also appear in other theories, 
such has dilatonic theories derived from the low-energy actions 
of string theory \cite{horowitz,caizhang,caietal}.
We do not dwell on these very interesting topics here.

\subsection{(2+1)- and (1+1)-dimensional black holes}

The black holes in 3+1-dimensions, i.e., in four-dimensions (4D),
presented in section 2, are also solutions in (2+1)-dimensions (3D) 
\cite{lemos1} and in (1+1)-dimensions (2D) \cite{lemos2}. In fact,
since the $z$ and $\phi$ directions are Killing directions one can
perform a dimensional reduction in one direction, say the  $z$ direction to
get a 3D black hole, and in both $z$ and $\phi$ directions to get a
2D black hole. 

By dimensionally reducing the action (\ref{action1}) 
through one dimension one gets the following 3D action,
$S=\int d^3x \sqrt{-g} e^{-2\phi}
\left( R - 4 \omega \left( \partial \phi \right)^2 + 4 \lambda^2\right)\,,$
where
$g$ is the determinant of the 3D metric,
$R$ is the 3D curvature scalar,
$\phi$ is a scalar called the dilaton field,
which has come from the extra dimension,
$\lambda$ is a constant (related to $\Lambda$),
and $\omega=0$ in this reducing process, but can be made a free 
parameter in general.
This is a Brans-Dicke action in 3D.  Now, each different $\omega$ can
be viewed as yielding a different dilaton gravity theory.  For
instance, for $\omega=0$ it yields the above theory, equivalent in
certain ways to 4D general relativity \cite{lemos1}; for $\omega=-1$
one gets the simplest low energy string action \cite{call};  for
$\omega=\infty$ one obtains 3D general relativity 
and its BTZ black hole \cite{bana,henneaux}
(see \cite{cardoso4} for perturbation on this black hole and 
references therein, see \cite{brill} for multi-black hole solutions).  Each
different $\omega$ has a very rich and non-trivial structure of
solutions with many black holes, which can be considered on its own.
Here we do not attempt to comment on the whole set of solutions of the
different theories \cite{salemos3d1,salemos3d2}. For electric and 
magnetic rotating solutions see \cite{dias1,dias2}.  

Analogously, one gets on reduction the action (\ref{action1}) 
through two dimensions the following 2D action, 
$S=\int d^2x \sqrt{-g} e^{-2\phi}
\left( R - 4 \omega \left( \partial \phi \right)^2+ 4 \lambda^2 
  \right) \,,$
where now $\omega=-\frac12$ in this reducing process, 
but can be made again a free parameter in general. Thus 
$\omega=-\frac12$ yields the above 2D theory, equivalent in
certain ways to 4D general relativity \cite{lemos2};
for $\omega=-1$ one gets the simplest low energy string
action \cite{call}; for
$\omega=\infty$ one obtains the two dimensional analogue of general
relativity \cite{salemos2d1}.
We refer to the appropriate papers and citations in the following papers
\cite{salemos2d2}$^-$\cite{dias1}.

\subsection{Higher-dimensional black holes}

Higher dimensional black holes appear in several gravity theories 
and some can have an interpretation in 4D upon a suitable Kaluza-Klein 
dimensional reduction  (see \cite{lemosreview} for a review
in higher and lower dimensions). 
Black holes in dimensions higher than 4 in anti-de Sitter backgrounds
have been found \cite{birmingham,klemm2}. All these black holes, in low, in
4, and in higher dimensions are important in connection to the AdS/CFT
conjecture.  This conjectures states a correspondence between
supergravity (the low energy limit of string theory) on an anti-de
Sitter space and a conformal field theory on the boundary of that
space \cite{maldacena,witten}.  For instance, due to this AdS/CFT
duality, quasi-normal frequencies in the BTZ black hole spacetime
yield a prediction for the thermalization timescale in the dual
two-dimensional conformal field theory, which otherwise would be very
difficult to compute directly. If one has, e.g., a 10-dimensional type
IIB supergravity, compactified into a ${\rm BTZ} \times S^3 \times
T^4$ spacetime, the scalar field used to perturb the BTZ black hole,
can be seen as a type IIB dilaton which couples to a conformal field
theory operator $\cal{O}$. Now, the BTZ in the bulk corresponds
to a thermal sate in the boundary conformal field theory, and
thus the bulk scalar perturbation corresponds directly to a thermal
perturbation with nonzero $<\cal O>$ in the conformal field
theory. Similar arguments related to black hole thermodynamics in 4
and higher dimensions also hold \cite{birmingham,witten}.

\section{Conclusions}

We have reviewed the toroidal, cylindrical and planar black holes that
appear as solutions of general relativity with a negative cosmological
constant and study some of their properties. Other properties that
could be studied are the uncommon behavior of the luminosity
\cite{klemm3}, the propagation of vacuum polarized photons
\cite{cai}, and black hole pair production 
(see \cite{dias3,dias4} and references therein)
to name three. 

What are the uses of these black holes?

In astrophysics, as we know it,
little, apart some better understanding of the hoop
conjecture: formation of cylindrical black holes in anti-de Sitter 
spacetimes provide a direct counter-example to the conjecture. 

For general relativity, they yield new exact solutions that should be
added to the Kerr-Newman family.  The existence of topological black
holes enlarges the field of geometry in which general relativity and
related theories act. The beautiful pure geometric constructions of
the BTZ 3D black hole \cite{henneaux,brill} and of the constant
curvature 4D black holes \cite{aminneborg} are there to prove it.

In theories of elementary particles dealing with Planck scales, such
as string theory with its elementary strings and branes, a black hole
is itself considered as a phase state of elementary strings with about
one Planck mass obeying the laws of quantum mechanics.  In such
theories all types of black holes in any permitted dimension must be
important.  These black holes we have been discussing are certainly
important just above the Planck scales, where there are indications
that the vacuum is anti-de Sitter. They might be detected one day
through gravitational radiation, Hawking evaporation, or black hole
pair production.  In some instances, one can now calculate from first
principles the black hole entropy \cite{stromingervafa} (see
\cite{maldacena2,damour,peet} for reviews).  This leads to arguments in favor
of the resolution of the information paradox for certain black
holes. However, topological black holes in anti-de Sitter spacetimes
do not fit easily into those arguments
\cite{caldarelliklemm98,klemm}. As a last use, black holes in anti-de
Sitter spaces are at the center of the AdS/CFT conjecture mentioned
above.

The asymptotic anti-de Sitter black holes, like the asymptotically
flat ones, can appear in any scale from the Planck size to
astrophysical dimensions. This property of black holes 
is unique in a sense, all other objects,
either have a well defined scale, or a small range of scales 
(for a review on  the sizes of black holes see \cite{lemosreview2}).

We have tried to quote all the literature relevant and somehow 
connected to this subject. This was intended to be a six 
pages review, it has grown to more than the double, and it is now 
clear that the subject of topological black holes, not only
the ones mentioned here, 
deserve a thorough review.

\section*{Acknowledgments}
I thank Observat\'orio Nacional - Rio de Janeiro for hospitality, while 
part of this work was being written. 
This work was partially funded by FCT - Funda\c c\~ao para a Ci\^encia e 
Tecnologia through project ESO/PRO/1250/98.

\newpage
\topskip 0.5cm

\end{document}